\documentclass{article}
\usepackage{spconf,amsmath,graphicx,amssymb,subfigure}

\title{INDIVIDUALIZED HEAR-THROUGH FOR ACOUSTIC TRANSPARENCY USING PCA-BASED SOUND PRESSURE ESTIMATION AT THE EARDRUM}
\name{Wenyu Jin, Tim Schoof, Henning Schepker}
\address{Starkey Hearing Technologies, Eden Prairie, MN 55344, United States}
\begin{document}
\ninept
\def\baselinestretch{.903}\let\normalsize\small\normalsize
\maketitle
\begin{abstract}
The hear-through functionality on hearing devices, which allows hearing equivalent to the open-ear while providing the possibility to modify the sound pressure at the eardrum in a desired manner, has drawn great attention from researchers in recent years. To this end, the output of the device is processed by means of an equalization filter, such that the transfer function between external sound sources and the eardrum is equivalent for the open-ear and the aided condition with the device in the ear. To achieve an ideal performance, the equalization filter design assumes the exact knowledge of all the relevant acoustic transfer functions.  A particular challenge is the transfer function between the hearing device receiver and the eardrum, which is difficult to obtain in practice as it requires additional probe-tube measurements. In this work, we address this issue by proposing an individualized hear-through equalization filter design that leverages the measurement of the so-called secondary path to predict the sound pressure at the eardrum using a principle component analysis based estimator. Experimental results using real-ear measured transfer functions confirm that the proposed method achieves a good sound quality compared to the open-ear while outperforming filter designs that do not leverage the proposed estimator.
\end{abstract}
\section{Introduction}
\label{sec:intro}
In recent years, hearing devices with so-called hear-through functionalities that aim to provide awareness and means of communication to the wearer have become increasingly prevalent. However, recent studies demonstrated that only a few commercially available devices with hear-through functionalities succeed at achieving acoustic transparency, i.e., listening with the hearing device is the same as with the open-ear \cite{Denk2020aes,Schepker2020aes}. Acoustic transparency can be realized by means of sound pressure equalization algorithms, where a major challenge is the estimation of the sound pressure level that is generated by the hearing device at the individual eardrum.

Several hear-through algorithms that aim at achieving acoustic transparency have been proposed in the past \cite{Hoffmann2013aes,Raemo2014eusipco,Vaelimaeki2015spm,Denk2018ija,Denk2018itg,Gupta2019icassp,Fabry2019icassp,Schepker2021robust,schepker2021jasael}. In order to achieve an ideal sound pressure equalization, these algorithms assume knowledge, e.g., through acoustic measurements, of the acoustic transfer functions (ATFs) between the external sound field and the eardrum, as well as the ATF between the hearing device receiver and the eardrum. The sound pressure at the eardrum is subject to significant inter-individual variations due to personal ear canal acoustics. The best practice to obtain an accurate estimation of it requires so-called probe-tube measurements, where a small probe microphone is positioned close to the eardrum of the user to measure the sound pressure level \cite{Hellstrom1993}. However, the placement and measurement using a probe-tube is a delicate and time-consuming process.

Alternatively, the sound pressure at the eardrum can be estimated by using a microphone at the inner face of the hearing device. However, it is well known that the pressure at different locations inside the ear canal can vary substantially \cite{Stinson1985}. Hence, more recently, models that utilize electro-acoustic analogies to predict the pressure at the eardrum from a microphone at the inner face of the hearing device have received increased attention \cite{Hudde1999,SankowskyRothe2011,SankowskyRothe2015,Vogl2019}. While systems based on electro-acoustic analogies show a great potential to predict the sound pressure accurately up to 6--8 kHz \cite{SankowskyRothe2015,Vogl2019}, they are specific to each hearing aid design (e.g., depending on venting and acoustic transducer characteristics), and rely on delicate and time-consuming calibration routines \cite{Blau2010}. In this paper we present a novel supervised approach to estimate the sound pressure at the eardrum by forming a model based on a training set. The system requires a microphone at the inner face of the hearing aid mold  but does not rely on knowledge of the specific hearing aid design. 

The proposed individualized hear-through filter design uses measurements of the so-called secondary path, i.e., the acoustic impulse response between the hearing aid receiver and an inward-facing microphone at the in-ear earpiece, to predict the sound pressure at the eardrum. Specifically, the secondary path measurement is projected onto a lower dimensional space using principal component analysis (PCA) and transformed to obtain an estimate of the sound pressure at an individual's eardrum, similar to the method proposed in \cite{Fabry2020icassp} to estimate the primary path for feedforward active noise control. Results demonstrate that when this estimate is incorporated into the design of an individualized hear-through equalization filters for acoustic transparency, the proposed approach achieves good sound quality compared to the open-ear and has better performance than individualization based on using the secondary path itself.

\section{Problem Formulation}
\label{sec:2}
\begin{figure}[t]
\centerline{\includegraphics[width=0.73\linewidth]{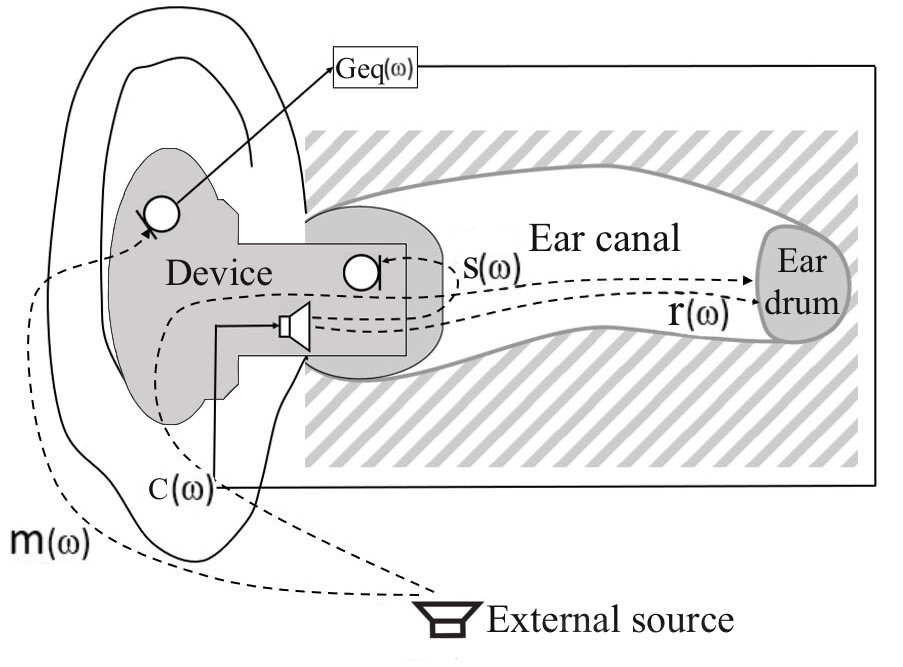}}
\caption{\label{fig:FIG2}\begin{footnotesize}The acoustic scenario for the hear-through filter design.\end{footnotesize}}
\end{figure}
The considered acoustic setup of the hearing device is depicted in Figure \ref{fig:FIG2}. To achieve acoustic transparency, the transfer function between external sound sources and the eardrum should be equivalent for the open ear and the aided ear (i.e., when the device is inserted and processing the sound).  In the open-ear case, the signal at the eardrum is the source signal filtered by the ATF to the eardrum of the open-ear $o(\omega)$. For the aided case with the device inserted in the ear, the signal at the eardrum is the superposition of a source signal filtered by the ATF of the direct path $c(\omega)$ (which is subject to occlusion) and the source signal filtered by the acoustic path to the external microphone $m(\omega)$, the equalization filter $G_{eq}(\omega)$ and the ATF between the inward-facing receiver and the
eardrum $r(\omega)$. It should be noted that all processing
delays are included in $r(\omega)$. Furthermore, we neglect the acoustic coupling between the receiver and the external microphone (i.e. acoustic feedback). To match the two cases, the equalization filter $G_{eq}(\omega)$ needs to be computed in the frequency domain such that
\begin{equation}
o(\omega)=c(\omega)+m(\omega)G_{eq}(\omega)r(\omega).
\label{eq:EQ7}
\end{equation}
The optimal equalization filter that achieves acoustic transparency in \eqref{eq:EQ7} is equal to
\begin{align}
    G_{eq}(\omega) = \frac{o(\omega)-c(\omega)}{m(\omega)r(\omega)},
\end{align}
and relies on the availability of all ATFs including $o(\omega)$, $c(\omega)$, $m(\omega)$, and $r(\omega)$. In the following we will present how to compute the optimal equalization filter $G_{eq}(\omega)$ using a least-squares-based cost function. 

\section{\label{sec:3} Individualized hear-through Filter Design}
In this section, we aim to incorporate a measured database of multiple sets of measurements of $o(\omega)$, $c(\omega)$, $m(\omega)$ in combination with a novel PCA-based estimator of the individual $r(\omega)$ to compute an individualized equalization filter that is based on the in-situ measurement of the so-called secondary path $s(\omega)$ between the hearing device receiver and the inward-facing microphone.
\subsection{\label{subsec:3:1} Equalization filter design}
The equalization filter in \eqref{eq:EQ7} aims at minimizing the difference between the aided and open-ear transfer function to the eardrum. This can be achieved by solving the following regularized least-squares (LS) optimization problem:
\begin{equation}
\underset{\mathbf{G_{eq}}}{\mbox{argmin}} \ \|(\mathbf{c}+\mathbf{D_{m}D_{r}G_{eq}})-\mathbf{o}\|_{2}^2+\mu\|\mathbf{G_{eq}}\|^{2}_{2},
\label{eq:EQ8}
\end{equation}
where $\mathbf{D_{m}}$, $\mathbf{D_{r}}$ are diagonal matrices containing the DFT coefficients of $m(\omega)$, $r(\omega)$, respectively. $\mathbf{c}$ and $\mathbf{o}$ are corresponding vector forms of $c(\omega)$ and $o(\omega)$. The Tikhonov regularization factor $\mu=0.001$ is considered in this work to prevent over-amplification of $\mathbf{G_{eq}}$. The optimal LS solution $\mathbf{G}_{eq}^{ls}$ is equal to
\begin{equation}
\mathbf{G}_{eq}^{ls}=(\mathbf{Y}^{H}\mathbf{Y}+\mu\mathbf{I})^{-1}\mathbf{Y}^{H}(\mathbf{o}-\mathbf{c}),
\label{eq:EQ9}
\end{equation}
where $\mathbf{Y}=\mathbf{D_{m}D_{r}}$, $\mathbf{I}$ is an identity matrix and $(\cdot)^{H}$ denotes the Hermitian transpose of a matrix. Similar to the frequency-domain filter in \eqref{eq:EQ9}, the time domain equalization filter  $\mathbf{g}_{eq}^{ls}$ of length $N_t$ can be computed as:
\begin{equation}
\mathbf{g}_{eq}^{ls}=(\mathbf{Y_D}^{H}\mathbf{Y_D}+\mu\mathbf{I})^{-1}\mathbf{Y_D}^{H}(\mathbf{o}-\mathbf{c}),
\label{eq:EQ10}
\end{equation}
where $\mathbf{Y_D}=\mathbf{D_{m}D_{r}Z_{D}\underline{F}}$ with $\mathbf{Z_{D}}$ denoting a diagonal matrix whose elements are the phase coefficients corresponding to a time shift $d_{proc}$ and
\begin{equation}
\mathbf{\underline{F}} = \mathbf{\underline{T}}^{N_f \times N_f} [\mathbf{I}^{N_t \times N_t};\mathbf{O}^{N_f-N_t \times N_t}].
\end{equation}
$\mathbf{\underline{T}}$ is the DFT matrix, $N_f$ is the FFT length, $\mathbf{I}$ represents an identity matrix and $\mathbf{O}$ is a matrix containing only zeros. The negative shift in time by $d_{proc}$ is considered to avoid potential acausality problems \cite{Denk2018itg,Fabry2019icassp,Schepker2021robust} and it is chosen to reflect the processing delay, which is set to 1.6 ms in this work. 

To facilitate the use of an estimator of the individual receiver-to-eardrum function \eqref{eq:EQ10} can be rewritten as: 
\begin{equation}
\mathbf{\bar{g}}_{eq}^{ls}=(\mathbf{\bar{Y}_D}^{H}\mathbf{\bar{Y}_D}+\mu\mathbf{I})^{-1}\mathbf{\bar{Y}_D}^{H}(\mathbf{o-\mathbf{c}}),
\label{eq:EQ11}
\end{equation}
where $\mathbf{\bar{Y}_D}=\mathbf{{D}_{m}\bar{D}_{r}Z_{D}\underline{F}}$. The diagonal elements of the matrix $\mathbf{\bar{D}_{r}}$ are the coefficients of an individual estimate of the transfer function between the hearing aid receiver and the eardrum, which will be obtained using an appropriate estimator in the following section. 
\subsection{Estimation of receiver-to-eardrum transfer function}
\label{subsec:3:2}
A simple and intuitive way of obtaining an estimate of $\mathbf{\bar{D}_{r}}$ is to measure the impulse response of the secondary path. However, the sound pressure at the position of the microphone may be substantially different from the sound pressure at the eardrum. Therefore, in this section, we present an estimation scheme to map the measured secondary paths at the inward-facing microphone in the ear canal to the receiver-to-eardrum responses. The first subsection introduces a least-square regression method that minimizes mean-squared error of the estimated eardrum response coefficients in the frequency domain. Subsequently, we propose an estimation method that benefits from the numerical robustness and efficiency of the PCA. Finally, we motivate a selection scheme that combines the two presented estimators so that a consistent estimation of the eardrum sound pressure can be achieved across frequencies.
\subsubsection{\label{subsubsec:3:2:1} Linear least-squares regression}
Let $\mathcal{M}=\{\mathbf{s_{j}}, \mathbf{r_{j}} \in \mathbb{C}^{\frac{N_f}{2}+1} | j=1,\ldots,J\}$ be a set of DFT coefficients of measured transfer functions $s_{j}(\omega)$ and $r_{j}(\omega)$ used to compute a least-squares based estimate of the receiver-to-eardrum transfer function during a training stage, which can be conducted offline. The optimal filter $\mathbf{g_{LS}}$ should minimize the difference between the estimated receiver-to-eardrum transfer functions to the measured counterparts via linear mapping in the frequency domain, i.e., the following least-squares cost function
\begin{equation}
E(\mathbf{g_{LS}})=\|\mathbf{D_{s}}\mathbf{g_{LS}}-\mathbf{d_{r}}\|^{2}_{2}+\mu\|\mathbf{g_{LS}}\|^{2}_{2},\label{eq:EQ1}
\end{equation}
where $\mathbf{D_{s}}$ ($J(\frac{N_f}{2}+1) \times J(\frac{N_f}{2}+1)$) are diagonal matrices containing the DFT coefficients of all measured secondary path responses $s_{j}(\omega)$ and $\mathbf{d_{r}}$ ($J(\frac{N_f}{2}+1) \times 1$) is the stacked vectors containing the DFT coefficients of $r_{j}(\omega)$. The optimum with respect to $\mathbf{g_{LS}}$ is given by
\begin{equation}
\mathbf{\hat{g}_{LS}}=(\mathbf{D_{s}}^{H}\mathbf{D_{s}}+\mu \mathbf{I})^{-1}\mathbf{D_{s}}^{H}\mathbf{d_{r}}.
\end{equation}
Using the estimated filter $\mathbf{\hat{g}_{LS}}$ from the training stage, the individual receiver-to-eardrum path $\mathbf{\hat{r}_{LS}}$ is estimated from a measurement of the individual secondary path $\mathbf{s}$ during run-time as follows: 
\begin{equation}
\mathbf{\hat{r}_{LS}}=\mathbf{s} \odot \mathbf{\hat{g}_{LS}}, \label{eq:EQ5}
\end{equation}
where $\odot$ denotes element-wise product.
\subsubsection{\label{subsubsec:3:2:2} PCA-based estimation}
As it can be seen from \eqref{eq:EQ1}, a direct linear mapping of the complex frequency domain vectors $\mathbf{s_{j}}, \mathbf{r_{j}}$ is possible but would require a large set of training data if FFT length is large. In this section, an PCA-based estimator is designed for the individual $\mathbf{r}$ based on features of a measured individual secondary path $\mathbf{s}$.

PCA is commonly used for dimensionality reduction by projecting each data point onto only the first few principal components to obtain lower-dimensional data that facilitates to avoid over-fitting \cite{Steward93}. By conducting PCA, we extract the first $K$ principal components $U_{s,k},U_{r,k} \in \mathbb{C}^{\frac{N_f}{2}+1}$ of the set of complex frequency domain vectors $\mathbf{s_{j}}$ and $\mathbf{r_{j}}$, respectively. The receiver-to-eardrum path principal components matrix is defined as
\begin{equation}
\mathbf{U}_{r}= [U_{r,1},U_{r,2},\ldots,U_{r,K}].
\end{equation}
Let $\mathbf{\bar{r}}$ be the ensemble average of $\mathbf{r_{j}}$: $\mathbf{\bar{r}}=\sum_{j \in \mathcal{M}} \mathbf{r_{j}}/J$. To obtain the complex gain vectors $\mathbf{g}_{r,j}$ that minimize the Euclidean distance between the reconstructed frequency domain vectors
\begin{equation}
\mathbf{\hat{r}}_{j}= \mathbf{\bar{r}}+\mathbf{U}_{r}\mathbf{g}_{r,j}
\end{equation}
and the true frequency domain vectors $\mathbf{r}_{j}$, we utilize the orthonormality of the principal components and get
\begin{equation}
\mathbf{g}_{r,j}=\mathbf{U}_{r}^{H}(\mathbf{r}_{j}-\mathbf{\bar{r}}).
\end{equation}
Similarly we obtain the gains $\mathbf{g}_{s,j}$ for the secondary path.

After converting frequency domain coefficients into the principal component domain, the problem is to find a linear map $\mathbf{A} \in \mathbb{C}^{K \times K}$ that projects the secondary path gain vectors onto the receiver-to-eardrum gain vectors. The following cost function is defined:
\begin{equation}
E(\mathbf{A})= \sum_{j \in \mathcal{M}}  \|\mathbf{\tilde{g}}_{r,j}-\mathbf{A}\mathbf{\tilde{g}}_{s,j}\|^{2},\label{eq:EQ2}
\end{equation}
with $\mathbf{\tilde{g}}=\mathbf{g}-\mathbf{\bar{g}}$ and $\mathbf{\bar{g}}$ denotes the ensemble average. The linear map allows us to estimate the receiver-to-eardrum path gain vector based on the secondary path gain vector. To minimize $E(\mathbf{A})$, we have
\begin{equation}
\mathbf{\hat{A}}= \underset{\mathbf{A}}{\mbox{argmin}} \ E(\mathbf{A})=\sum_{j \in \mathcal{M}} \mathbf{\tilde{g}}_{r,j}\mathbf{\tilde{g}}_{s,j}^{H} (\sum_{j \in \mathcal{M}} \mathbf{\tilde{g}}_{s,j}\mathbf{\tilde{g}}_{s,j}^{H})^{-1}.
\label{eq:EQ3}
\end{equation}

The above-demonstrated steps are the training stage for individual receiver-to-eardrum response estimation based on a training set $\mathbf{s_{j}}, \mathbf{r_{j}} \in \mathcal{M}$. After measuring the individual secondary path $\mathbf{s}$ at runtime, the gain
vector for the secondary path can be calculated as follows:
\begin{equation}
\mathbf{g}_{s}=\mathbf{U}_{s}^{H}(\mathbf{s}-\mathbf{\bar{s}}),
\end{equation}
where $\mathbf{\bar{s}}$ is the ensemble average of $\mathbf{s}_{j}$ from the training stage. We can then obtain an estimate for $\mathbf{g}_{r}$ by
\begin{equation}
\mathbf{\hat{g}}_{r}=\mathbf{\bar{g}}_{r}+\mathbf{\hat{A}}\mathbf{g}_{s}
\end{equation}
and finally an estimate $\mathbf{\hat{r}_{pca}}$ for $\mathbf{r}$ in the frequency domain with the ensemble average $\mathbf{\bar{r}}$ and $\mathbf{U}_{r}$ from the training stage
\begin{equation}
\mathbf{\hat{r}_{pca}}=\mathbf{\bar{r}}+\mathbf{U}_{r}\mathbf{\hat{g}}_{r}. \label{eq:EQ6} 
\end{equation}
\subsubsection{\label{subsubsec:3:2:3}Estimate selection}
Two linear estimators of individual receiver-to-eardrum path transfer functions are presented in Sec.~\ref{subsubsec:3:2:1} and Sec.~\ref{subsubsec:3:2:2}, respectively. For the PCA-based estimation method, it is suggested to extract frequency regions that are affected by deterministic changes of ear canal characteristics so that features of the measured individual secondary path $\mathbf{s}$ can be better utilized. It is shown in \cite{SankowskyRothe2015} that $\mathbf{s_{j}}$ and $\mathbf{r_{j}}$ are less differentiated at low frequency regions, which can be intuitively explained by the long wavelength. Therefore, we propose to combine both the low and high frequencies into the final receiver-to-eardrum path estimate $\hat{r}(\omega)$ and synthesize the two estimators as follows:
\begin{equation}
\hat{r}(\omega)=\left\{
\begin{aligned}
\hat{r}_{LS}(\omega), \mbox{for} \ \omega \leq \omega^{\prime} \\
\hat{r}_{pca}(\omega), \mbox{for} \ \omega > \omega^{\prime}
\end{aligned}
\right.
\label{eq:EQ6_1}
\end{equation}
where $\omega^{\prime}$ denotes the split frequency that separates the two estimators at low and high frequencies. In this work, we empirically select $\omega^{\prime}=1.5$ kHz. The effectiveness of this selection scheme is confirmed in Sec. 4. For practical considerations, we also propose to apply rectangular frequency domain
windows $Q_{LS}(\omega)$ and $Q_{pca}(\omega)$ to 
extract corresponding frequency regions of  the training set $\mathbf{s_{j}}, \mathbf{r_{j}} \in \mathcal{M}$ and the measured secondary path $\mathbf{s}$ at run time. The transition frequency for the low-pass $Q_{LS}(\omega)$ is $\omega^{\prime}$ and the pass-band for $Q_{pca}(\omega)$ is from $\omega^{\prime}$ to 8 kHz.

\section{\label{sec:4}System Evaluation}

In this section, we evaluate the effectiveness of incorporating the proposed receiver-to-eardrum transfer function estimator into the individualized equalization filter design that aims to achieve acoustic transparency. First the acoustic setup and the database of measurements are introduced. Second, we present results that demonstrate the estimation accuracy of the proposed receiver-to-eardrum transfer function estimator. Finally we present results of a perceptual evaluation of the proposed individualized equalization filter utilizing the multi stimulus hidden reference and anchor (MUSHRA) framework.

\subsection{\label{subsec:4:1} The in-the-ear hearing aid prototype database}
ATFs were measured in the right ear for 18 subjects using an occluding in-the-ear hearing aid prototype (as shown in Fig.~\ref{fig:FIG2}) that features one in-ear Sonion P8 MEMS microphone, one external Sonion P8 MEMS microphone and one in-ear Knowles RAB balanced-armature receiver. All acoustic transducers were connected to a PC via a soundcard and a custom amplifier box using a sampling rate of 40 kHz. The external sound sources were placed on a circle at 1.5 m distance around the position of the listener. All relevant transfer functions to the eardrum were measured using an audiological probe tube connected to an Etymotics ER7C microphone inserted into the ear canal using a fixed insertion depth from the inter-tragal notch (i.e. 28 mm for women, 30 mm for men \cite{Pumford01}). To reduce the effect of placement of the microphone away from the eardrum, e.g., quarter-notch wavelength notches \cite{Searchfield97}, the probe-tube microphone responses were corrected using the eardrum sound pressure prediction method in \cite{SankowskyRothe2011}. Each measurement was repeated for three reinsertion trials of the device for each of the subjects. Overall, a total of 54 transfer function sets were considered for evaluation. We averaged the measured responses over reinsertion trials and used the average data for each of the 18 subjects in our assessment. When evaluating the performance we used a repetitive leave-one-out cross-validation approach, i.e., we used data from 17 subjects for training and evaluated the performance for the remaining subject. 

\subsection{\label{subsec:4:2}Estimation of receiver-to-eardrum transfer function}

To evaluate the performance of the eardrum sound pressure estimators, we used the following measure to quantify the dB level error between the estimated eardrum sound pressure $\hat{r}(\omega)_{j}$ and the ground-truth ${r}(\omega)_{j}$ for the $j$th subject across frequencies:
\begin{equation}
\epsilon(\omega)_{j}=10\log(|{r}(\omega)_{j}|^2)-10\log(|\hat{r}(\omega)_{j}|^2).
\end{equation}
The estimation performance for all 18 subjects using the proposed method that combines the PCA-based estimator (in which we extracted the first $K=12$ principal components) and LS regression with the split frequency $\omega^{\prime}=1.5$ kHz are demonstrated in Fig. 2. The bold line represents the mean estimation error and the shaded area covers the standard deviation across subjects. An estimation accuracy within $\pm 5$\,dB for frequencies up to 6 kHz can be achieved, which is comparable to the systems based on electro-acoustic analogies \cite{SankowskyRothe2015,Vogl2019} but does not require delicate calibrations on specific earpiece designs. Fig.~\ref{fig:FIG6} presents error regions across frequencies for the two estimation schemes separately. It can be seen that a noticeably improved estimation performance can be achieved by the PCA–based approach at mid- and high- frequencies compared to the LS-based method. However, it does not perform as consistently as the LS-based method at the low frequency range with larger error variances, which justifies the estimate selection method proposed in Sec. \ref{subsubsec:3:2:3}. 
\begin{figure}
\centering
\includegraphics[width=0.84\linewidth]{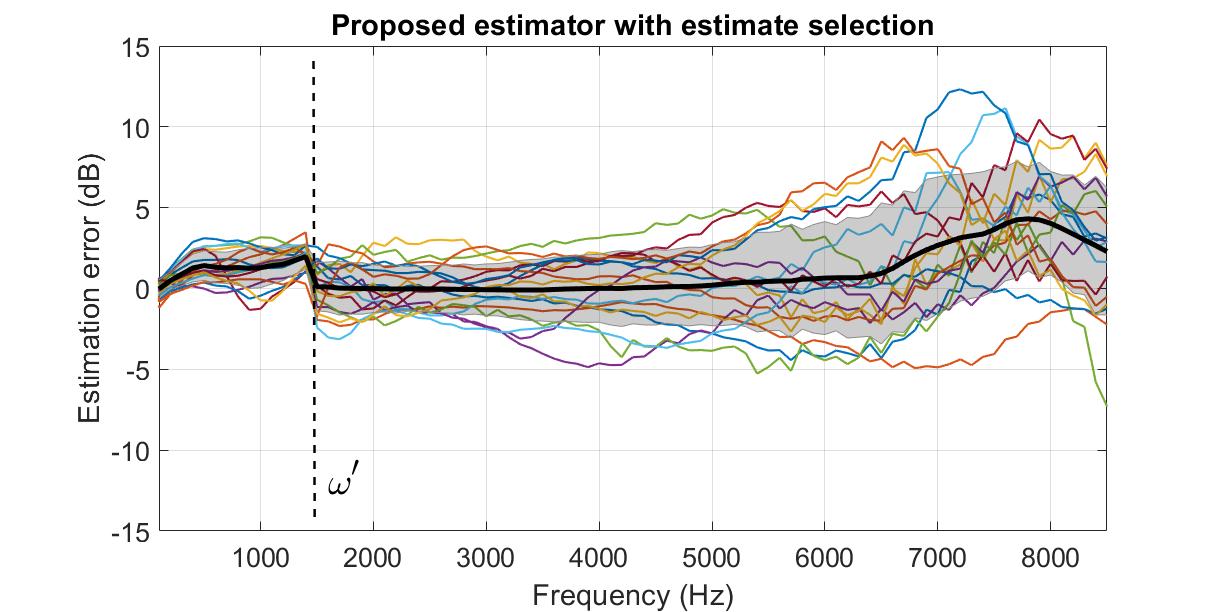}
\vspace{-0.2cm}
\label{fig:FIG5}
\caption{Estimation error of the proposed combined receiver-to-eardrum transfer function estimator for 18 subjects.}
\end{figure}
\begin{figure}[t]
\centerline{\includegraphics[width=0.84\linewidth]{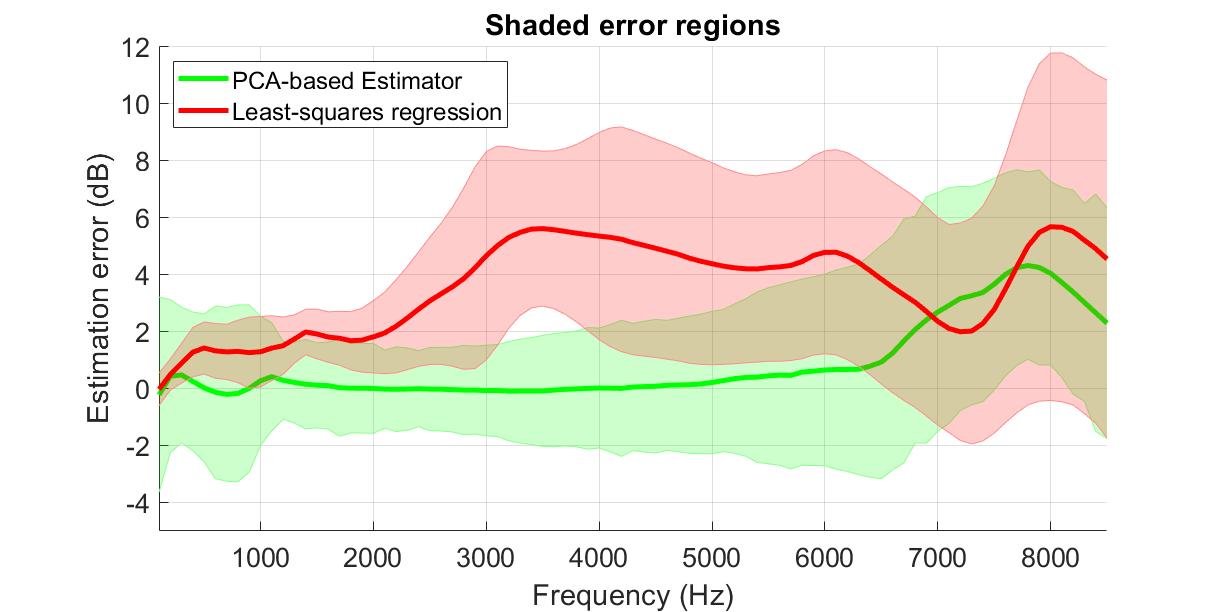}}
\vspace{-0.2cm}
\caption{\label{fig:FIG6}Average estimation error (bold lines) for the least-squares based estimator and PCA-based estimator across 18 subjects. The shaded areas indicate the standard deviation of the estimator error.}
\end{figure}

\subsection{\label{subsec:4:3} Individualized equalization filter}
By leveraging the estimate of the receiver-to-eardrum transfer function from Eq.~\eqref{eq:EQ6_1}, we compared the proposed individualized equalization ("IDV-PCA") in Eq.~\eqref{eq:EQ11} against the following processing conditions: 1) Open-ear without the device inserted; 2) Occluded ear with the devices switched off; 3) Perfect equalization, which denotes the individual equalization with the prior knowledge of all relevant ATFs; 4) "IDV-SP", which denotes an alternative individualization by directly replacing the diagonal elements of $\mathbf{\bar{D}_{r}}$ in Eq.~\eqref{eq:EQ11} with the coefficients of the secondary path $s_{j}(\omega)$; 5) "GLS", which denotes the average equalization using the least-squares method and leave-one-out cross-validated filter. This is achieved by replacing relevant diagonal matrices and vectors in \eqref{eq:EQ10} with stacked diagonal matrices and vectors that contain the DFT coefficients of relevant ATFs for all training subjects from the database. In this evaluation study, we chose time-domain filter length $N_{t}=64$.

\begin{figure}[t]
\centering  
\subfigure{
\label{Fig.sub.1}
\includegraphics[width=.48\linewidth]{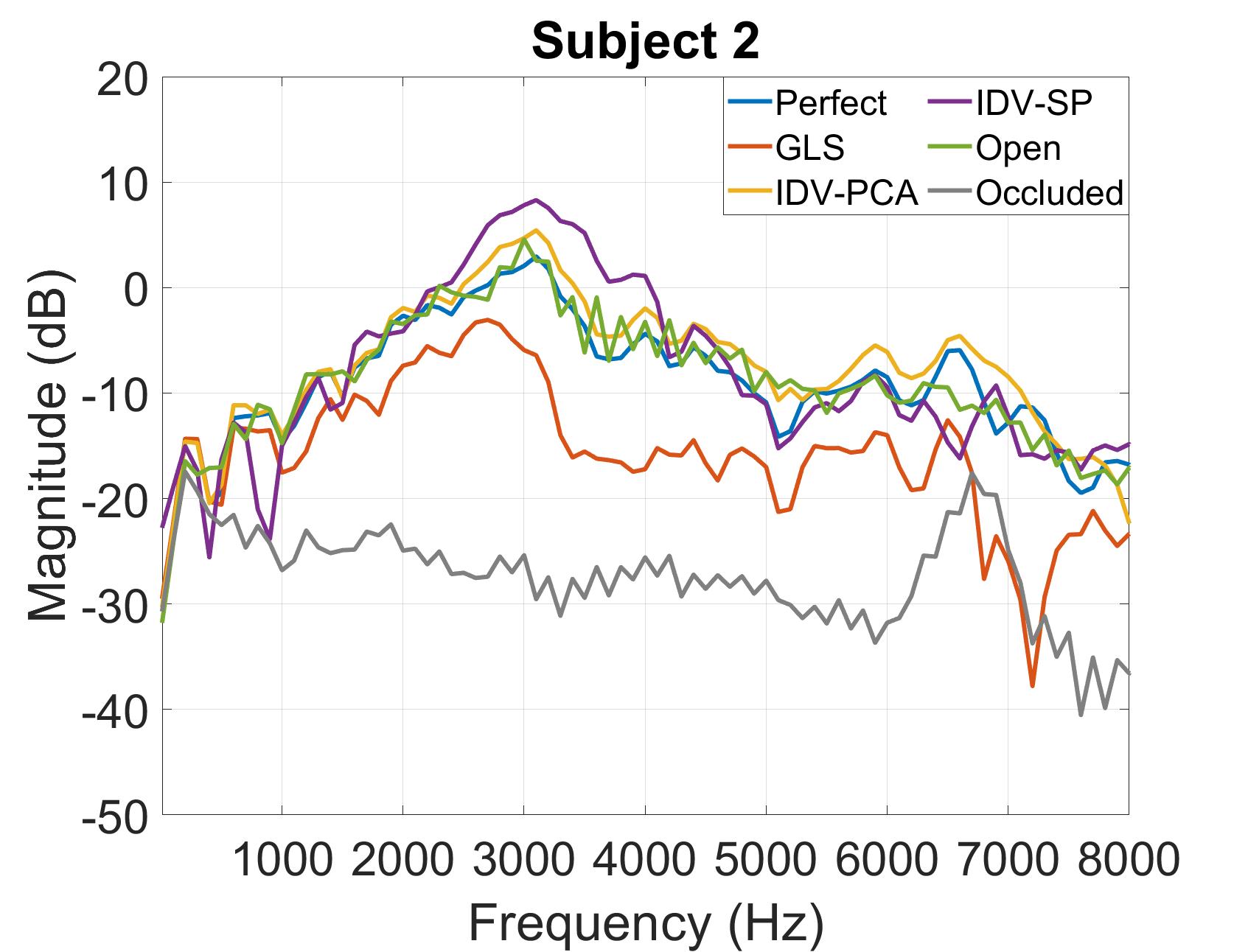}}\subfigure{
\label{Fig.sub.2}
\hfill
\includegraphics[width=.48\linewidth]{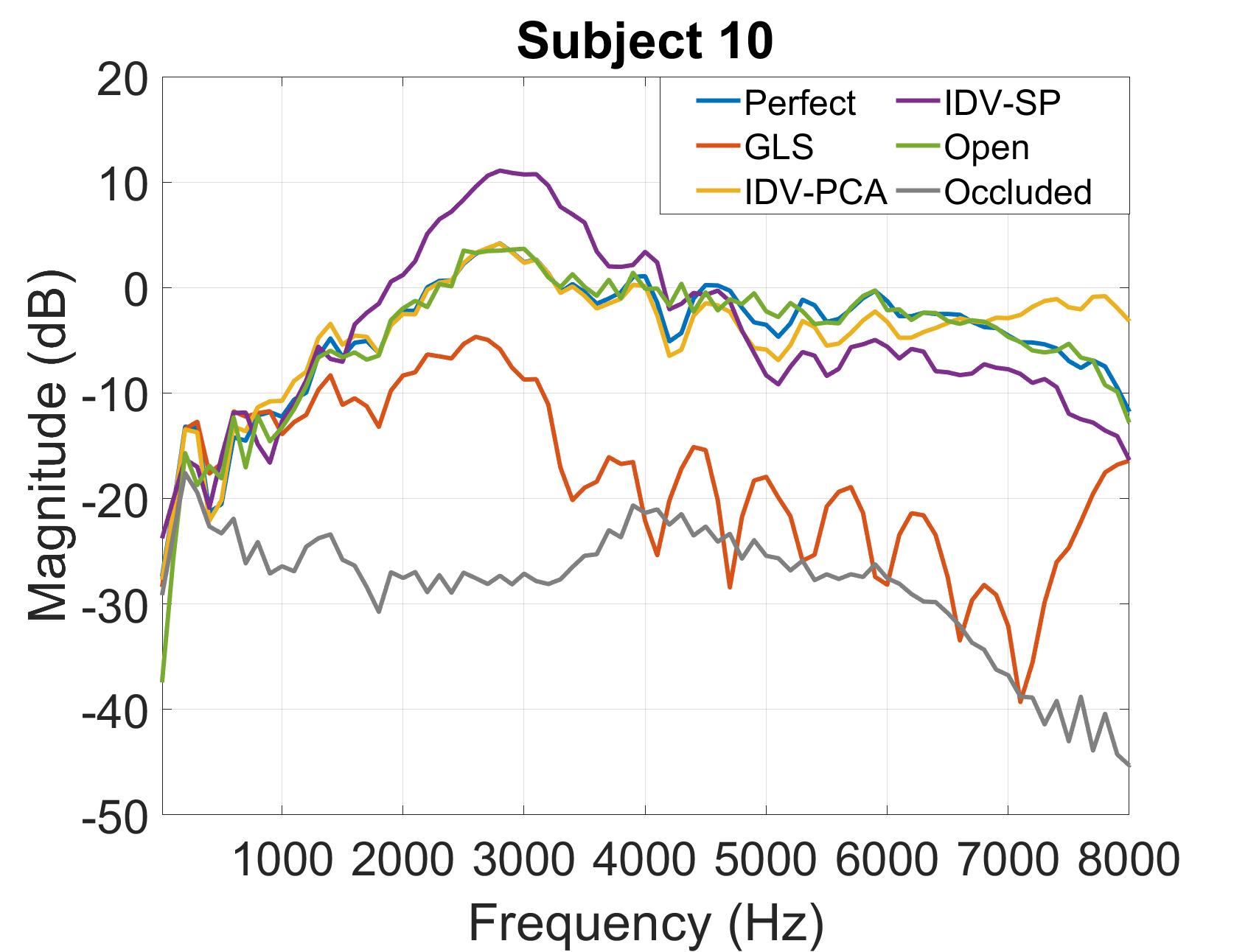}}
\vspace{-0.2cm}
\caption{Comparison of various hear-through equalization processing schemes for two exemplary subjects.}
\label{fig:FIG7}
\end{figure}

Figure \ref{fig:FIG7} shows magnitude responses for two exemplary subjects from the database. As expected, the best-performing system is the perfect equalization that assumes all ATFs to be known. Furthermore, the proposed IDV-PCA method provides a close match to the open-ear up to 6 kHz and consistently achieves a better hear-through performance compared to all remaining equalization filters. 

To investigate how well the proposed equalization filter reproduces the open-ear perception, we conducted a headphone-based MUSHRA listening test \cite{MUSHRA} with 12 self-reported normal-hearing listeners. For each listener, we randomly selected the ATFs from 3 subjects from the database to simulate the signals. Three 10-sec stimulus signals were considered, including  female speech, jazz music and street noise and where presented over a pre-equalized Beyerdynamic DT-770 headphones and played back at 70 dBSPL. Subjects were asked to rate the perceived sound quality in terms of naturalness in comparison to the provided open-ear reference samples. 

\begin{figure}[t]
\centerline{\includegraphics[width=.80\linewidth]{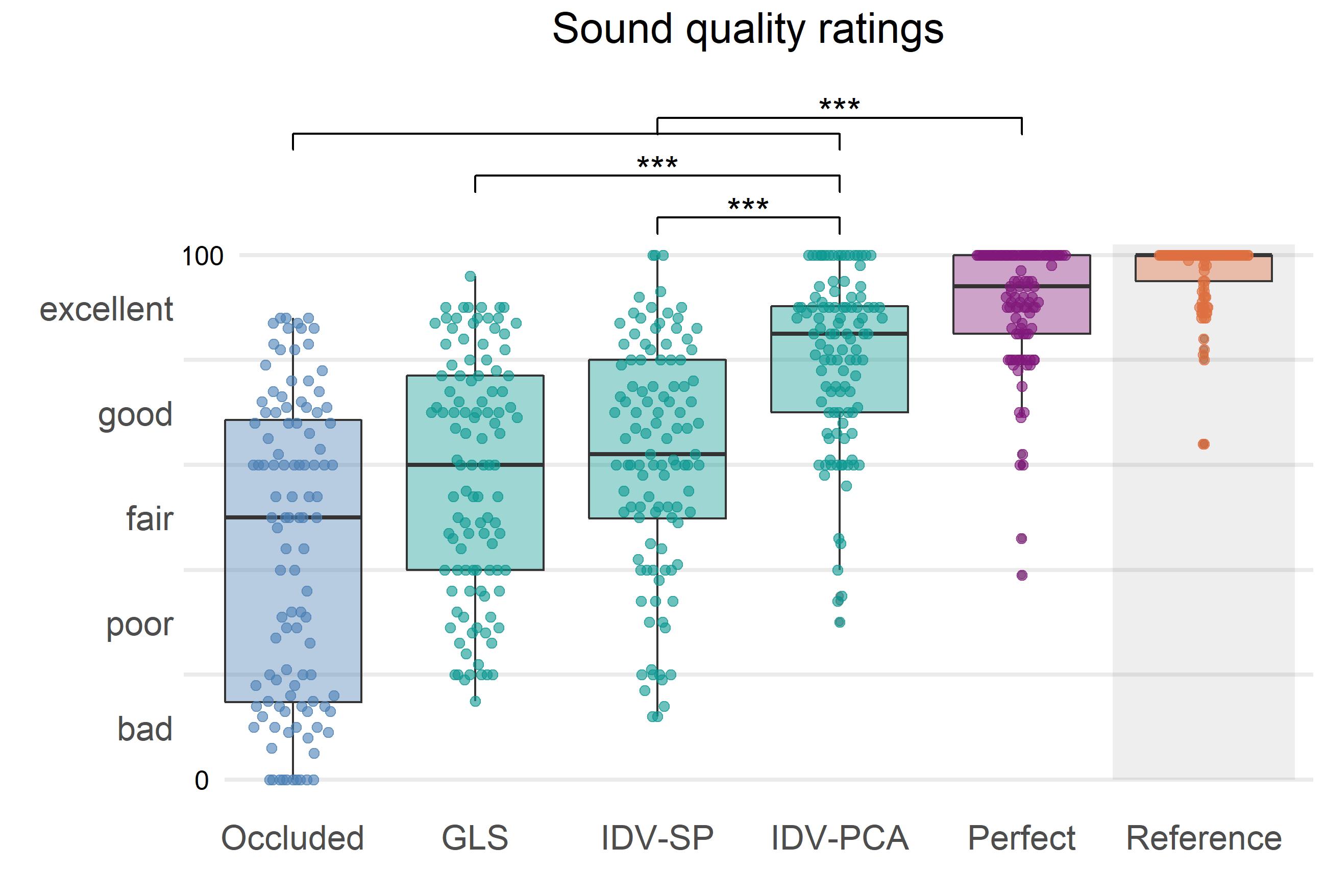}}
\vspace{-0.2cm}
\caption{\label{fig:FIG8}Sound quality ratings of the MUSHRA study evaluating the hear-through filters (*** indicates p $<$ 0.001).}
\end{figure}

Fig.~\ref{fig:FIG8} shows the results of the MUSHRA test. The results show that most subjects were able to identify the hidden reference signals. The remaining five conditions were analyzed using a linear mixed effects model \cite{lmmgui}. The results showed that the occluded ear was rated worst, and as expected from Figure \ref{fig:FIG7}, the highest ratings were obtained for the perfect equalization. Considering the non-perfect equalization filters, the proposed IDV-PCA approach resulted in the highest ratings ($p<0.001$ compared to GLS and IDV-SP). This shows the importance of incorporating information on the individual eardrum and validates the proposed PCA-based sound pressure estimation approach. 

\section{Conclusion}
\label{sec:conclusion}
In this work, we propose an individualized hear-through equalization filter design by leveraging an estimate of sound pressures at the eardrum. We present a PCA-based estimation approach to predict the receiver-to-eardrum transfer function using in-situ measurements of the secondary path at the in-ear microphone. By combining this with a least-squares-based estimator at low frequencies, the proposed method obtains an estimation accuracy within $\pm 5$\,dB for frequencies up to 6 kHz, which facilitates the proposed individualized equalization filter to deliver a good and natural sound quality compared to the open-ear while at the same time outperforming using the secondary path itself. A MUSHRA listening test was conducted and results confirm the effectiveness of the proposed approach.


\bibliographystyle{IEEEbib}
\bibliography{strings}

\end{document}